\documentclass[twocolumn,showpacs,preprintnumbers,amsmath,amssymb,superscriptaddress]{revtex4}

\usepackage{graphicx}
\usepackage{dcolumn}
\usepackage{bm}

\begin{document}

\preprint{APS/123-QED}

\title{Vortices and the entrainment transition in the 2D Kuramoto model}
\author{Tony E. Lee}
\affiliation{Department of Physics, California Institute of Technology, Pasadena, CA 91125}
\author{Heywood Tam}
\affiliation{Department of Physics, California Institute of Technology, Pasadena, CA 91125}
\author{G. Refael}
\affiliation{Department of Physics, California Institute of Technology, Pasadena, CA 91125}
\author{Jeffrey L. Rogers}
\affiliation{Control and Dynamical Systems, California Institute of Technology, Pasadena, CA 91125}
\author{M. C. Cross}
\affiliation{Department of Physics, California Institute of Technology, Pasadena, CA 91125}

\date{\today}

\begin{abstract}
We study synchronization in the two-dimensional lattice of coupled phase oscillators with random intrinsic frequencies. When the coupling $K$ is larger than a threshold $K_E$, there is a macroscopic cluster of frequency-synchronized oscillators. We explain why the macroscopic cluster disappears at $K_E$. We view the system in terms of vortices, since cluster boundaries are delineated by the motion of these topological defects. In the entrained phase ($K>K_E$), vortices move in fixed paths around clusters, while in the unentrained phase ($K<K_E$), vortices sometimes wander off. These deviant vortices are responsible for the disappearance of the macroscopic cluster. The regularity of vortex motion is determined by whether clusters behave as single effective oscillators. The unentrained phase is also characterized by time-dependent cluster structure and the presence of chaos. Thus, the entrainment transition is actually an order-chaos transition. We present an analytical argument for the scaling $K_E\sim K_L$ for small lattices, where $K_L$ is the threshold for phase-locking. By also deriving the scaling $K_L\sim\log N$, we thus show that $K_E\sim\log N$ for small $N$, in agreement with numerics. In addition, we show how to use the linearized model to predict where vortices are generated.

\end{abstract}

\maketitle

\section{\label{sec:level1}Introduction}

Collective behavior of coupled oscillators is found in many areas of science \cite{pikovsky01,acebron05}. Examples include Josephson junctions \cite{wiesenfeld96}, lasers \cite{silber93}, neural networks \cite{varela01}, chemical oscillators \cite{kuramoto84}, and nanomechanical resonators \cite{cross04}. Recently, there has been much interest in studying coupled oscillators on complex networks, motivated by biological and social networks \cite{strogatz01,dorogovtsev08,gomez07,mori10}.

This field is an interesting marriage of statistical physics and nonlinear science, because these non-equilibrium systems may exhibit phase transitions. A common approach is to consider populations of oscillators with random intrinsic frequencies. The coupling between oscillators acts against the frequency disorder to synchronize the oscillators.  The original Kuramoto model considered oscillators with all-to-all coupling \cite{kuramoto84}. The model has also been studied on low-dimensional lattices with local interactions \cite{sakaguchi87,daido88,strogatz88a,strogatz88b,hong05} and long-range interactions \cite{rogers96,marodi02,chowdhury10}.

Two synchronization transitions occur on low-dimensional models as the coupling $K$ changes. The entrainment transition at $K_E$ denotes the onset of macroscopic entrainment, when there is a cluster of frequency-synchronized oscillators on the order of the system size. There is also a transition at $K_L$ to the phase-locked state, in which all oscillators evolve with the same frequency. 

In the tradition of statistical physics, the main question is whether or not the entrainment transition exists, i.e., whether $K_E$ is finite, in the limit of infinite system size ($N\rightarrow\infty$). Simulations indicate that in the case of local interactions, it exists only in dimension $d\geq 3$, meaning that the lower critical dimension is 2 \cite{hong05}. Ideally, one would show this analytically, but it is difficult due to the presence of nonlinearity and disorder. It has been proven that macroscopic entrainment does not exist in $d=1$ in an infinite system \cite{strogatz88a,daido88}. Various heuristic arguments indicate that it exists only in $d\geq 3$ \cite{sakaguchi87,daido88,ostborn09}. However, there has been no clear explanation of exactly what happens at the entrainment transition.

In this paper, we examine the entrainment transition in the 2D model and elucidate why and how it happens in a finite system with local interactions. Surprisingly, there are several differences between the entrained phase ($K>K_E$) and the unentrained phase ($K<K_E$) besides the presence of the macroscopic cluster. The entrained phase is characterized by time-independent cluster structure, while cluster boundaries continually change in the unentrained phase. In fact, the entrained phase is not chaotic while the unentrained phase is chaotic, meaning that the entrainment transition is actually an order-chaos transition.

It is convenient to view the system in terms of vortices, since the boundaries of frequency clusters are delineated by vortex paths. In the entrained phase, vortices move in fixed paths around clusters, while in the unentrained phase, vortices sometimes wander off. These deviant vortices are responsible for the system-wide detrainment.

In an $N\times N$ lattice, the nature of the entrainment transition is different in small ($N\leq 50$) and large ($N\geq 100$) lattices. In small lattices, the transition is determined by a cluster made by a vortex pair, while in large lattices, the transition is due to clusters made by single vortices. We focus on small lattices and provide analytical arguments for the scalings $K_E\sim K_L$ and $K_L\sim\log N$. Hence, we derive the scaling $K_E\sim\log N$ for small $N$, in agreement with numerics. We also show how to use the linearized model to predict where vortices are created.

Previous works have touched on certain aspects of our results. Topological defects have been observed in 2D lattices of chaotic oscillators (where individual oscillators are intrinsically chaotic) \cite{davidsen02}. Chaos and phase slips have been observed in the 1D Kuramoto model \cite{zheng98}. Frequency clusters have been studied also in the 1D Kuramoto model \cite{kogan09,lee09}. This paper explains how all these ideas are related to each other and to the entrainment transition.

The outline of the paper is as follows. In Sec. II, we review the 2D model. In Sec. III, we summarize qualitatively its complex phenomenology. Then we present analytical results on cluster stability in Sec. IV. We study phase-locking in Sec. V and conclude in Sec. VI.

\section{\label{sec:level2}Model}

We consider the $N\times N$ two-dimensional lattice of oscillators with nearest-neighbor interactions and periodic boundary conditions:
\begin{eqnarray}
\label{eq:eom2d}
\dot{\theta}_{ij}&=&\omega_{ij}+ K[\sin(\theta_{i-1j}-\theta_{ij})+\sin(\theta_{i+1j}-\theta_{ij}) \nonumber\\
&& \quad\quad +\sin(\theta_{ij-1}-\theta_{ij})+\sin(\theta_{ij+1}-\theta_{ij})] \nonumber\\
&& \quad\quad\quad\quad\quad\quad i,j=1,\ldots,N
\end{eqnarray}
The intrinsic frequencies $\omega_{ij}$ are Gaussian distributed with zero mean and unit variance. We assume, without loss of generality, that the average of $\omega$ in a given realization is zero.

We also consider the linearized version:
\begin{eqnarray}
\dot{\theta}_{ij}&=&\omega_{ij}+ K[(\theta_{i-1j}-\theta_{ij})+(\theta_{i+1j}-\theta_{ij}) \nonumber\\
&& \quad\quad +(\theta_{ij-1}-\theta_{ij})+(\theta_{ij+1}-\theta_{ij})]
\label{eq:eom2dlin}
\end{eqnarray}
This approximation is sometimes useful, because for large enough $K$ and finite $N$, the lattice has small phase gradients and is thus in the linear regime. The obvious advantage of the linearized model is that it is straightforward to solve via discrete Fourier transforms. With the definitions
$\tilde{\theta}_{kl}=\sum_{mn}\theta_{mn}e^{-i\frac{2\pi}{N}(km+ln)}$ and
$\tilde{\omega}_{kl}=\sum_{mn}\omega_{mn}e^{-i\frac{2\pi}{N}(km+ln)}$, each $\tilde{\theta}_{kl}$ satisfies
\begin{eqnarray}\label{eq:eom_fourier}
\frac{d\tilde{\theta}_{kl}}{dt}=\tilde{\omega}_{kl}-2K(2-\cos{\frac{2\pi k}{N}}-\cos{\frac{2\pi l}{N}})\tilde{\theta}_{kl} \;.
\end{eqnarray}
Each Fourier component decays exponentially towards its steady-state value so that at steady state, $\theta_{mn}=\theta^{lin}_{mn}$, where
\begin{eqnarray}
\label{eq:2dss}
\theta^{lin}_{mn}&=&\frac{1}{N^2}\sum_{kl}\frac{\tilde{\omega}_{kl}e^{i\frac{2\pi}{N}(km+ln)}}{2K(2-\cos{\frac{2\pi k}{N}}-\cos{\frac{2\pi l}{N}})}
\end{eqnarray}
From Eq.~\eqref{eq:eom2dlin}, $\{\theta^{lin}_{mn}\}$ also satisfy:
\begin{eqnarray}
0=\omega_{ij}+ & K[(\theta^{lin}_{i-1j}-\theta^{lin}_{ij})+(\theta^{lin}_{i+1j}-\theta^{lin}_{ij}) \nonumber\\
& (\theta^{lin}_{ij-1}-\theta^{lin}_{ij})+(\theta^{lin}_{ij+1}-\theta^{lin}_{ij})]
\;, \label{eq:soln_2d}
\end{eqnarray}
The standard deviation of differences between neighboring phases can be calculated in the continuum approximation as \cite{hong05}:
\begin{eqnarray}
\label{eq:sddthlin}
\sigma_{\Delta\theta^{lin}}=\frac{1}{K}\sqrt{\frac{\log N}{4\pi}} \;.
\end{eqnarray}
Although this quantity diverges as $N\rightarrow\infty$, it does \emph{not} rule out the possibility of entrainment in 2D; there may be frequency order in the presence of large phase gradients, when the linear model is not applicable. Also, one might guess incorrectly from Eq.~\eqref{eq:sddthlin} that $K_E$ and $K_L$ scale as $(\log N)^{\frac{1}{2}}$.


The average frequency of an oscillator is defined as $\bar{\omega}=[\theta(t_0+T)-\theta(t_0)]/T$, where $t_0$ and $T$ are the transient and averaging times, respectively. An oscillator is said to be frequency-synchronized with its neighbor if their phase difference shifts by less than $\pi$ during the averaging time. In this paper, the numerical integration of Eqs. \eqref{eq:eom2d} was done using the Euler method with $t_0=T=10^4$, a time step of 0.02, and initial phases set to zero.

\section{\label{sec:level3}Entrained and unentrained phases}

The 2D Kuramoto model has a rich phenomenology. In this section, we qualitatively describe the behavior in order to motivate analytical calculations in later sections.

\subsection{\label{sec:level3a}Clusters}

\begin{figure}
\centering
\includegraphics[width=3.5 in, viewport=40 80 320 360, clip]{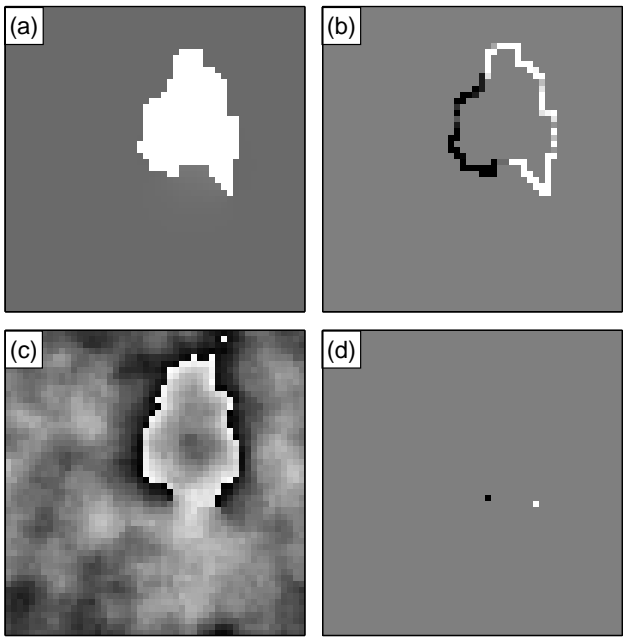}
\caption{\label{fig:lattice} Example of a $50\times 50$ lattice in the entrained phase. (a) Average frequency, showing the two frequency clusters. (b) Time-averaged vorticity, where black and white are opposite polarities. Vortex paths delineate the cluster boundaries. (c) Phase (mod $2\pi$) at a point in time, where black is 0 and white is $2\pi$. (d) The equivalent lattice of vortices at that time, where black and white are opposite polarities. In (a,c), each pixel represents an oscillator. In (b,d), each pixel represents a unit cell of four oscillators.}
\end{figure}

Consider what happens as $K$ changes. Above $K_L$, all oscillators have the same average frequency. Immediately below $K_L$, a small cluster of frequency-synchronized oscillators appears, while the rest of the system constitutes a macroscopic cluster [Fig.~\ref{fig:lattice}(a)]. As $K$ decreases further, more small clusters appear, and they generally maintain the same shape as $K$ decreases. Below $K_E$, there is no longer a macroscopic cluster.

There are three qualitative differences between the entrained and unentrained phases besides the presence of a macroscopic cluster. The first difference is the constancy of clusters. Above $K_E$, the cluster structure is time-independent after a sufficient transient time. However, below $K_E$, the cluster structure changes over time, so it is hard to say which oscillator is synchronized with which, since it depends on the values of $t_0$ and $T$. This time-dependence is surprising, since one would expect the unentrained phase to have small but well-defined clusters. (For example, the 1D chain with random $\omega$ and $K$ has a time-independent cluster structure, even in the unentrained phase \cite{kogan09,lee09}).

\begin{figure}
\centering
\includegraphics[width=3.5 in]{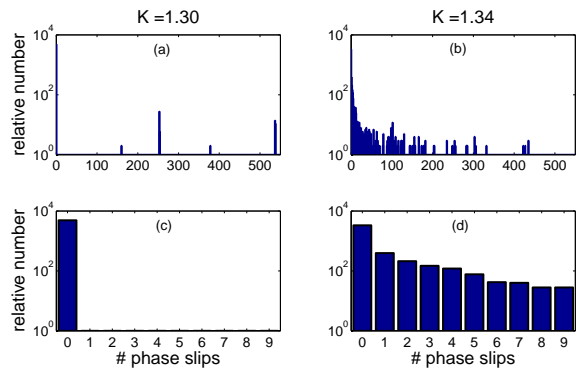}
\caption{\label{fig:dwavg} Distribution of the number of $2\pi$ phase slips between neighboring oscillators for a $50\times 50$ lattice. (a) In the entrained phase, neighbors have either zero or many phase slips during the averaging time. (b) In the unentrained phase, some neighbors have single or few phase slips. (c) and (d) are zoomed-in views of (a) and (b), respectively. Here, the entrainment transition happens at $K_E=1.32$.}
\end{figure}

\begin{figure}
\centering
\includegraphics[width=3.5 in]{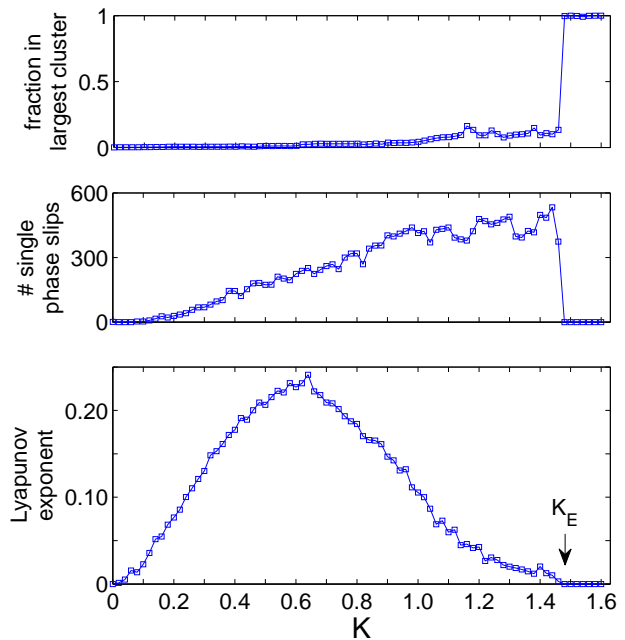}
\caption{\label{fig:qpsk} Size of the largest cluster, number of single phase slips, and largest Lyapunov exponent vs.\ coupling $K$ for a $50\times 50$ lattice. In this disorder realization, the entrainment transition happens at $K_E=1.48$, and phase locking happens at $K_L=1.58$. As $K$ decreases past $K_E$, the contraction of the largest cluster coincides with the appearance of single phase slips and chaos.}
\end{figure}

The second difference is in the distribution of average frequency differences between neighbors (Fig.~\ref{fig:dwavg}). Differences in average frequency are due to $2\pi$ phase slips. In the entrained phase, a pair of neighbors will experience either zero or many phase slips during the averaging time. But in the unentrained phase, some pairs experience only one or few phase slips. For a given disorder realization, as $K$ is decreased below $K_E$, there is a sudden drop in the size of the largest cluster and a sudden appearance of single $2\pi$ phase slips (Fig.~\ref{fig:qpsk}).

The third difference is that the entrained phase is not chaotic while the unentrained phase is. The largest Lyapunov exponent is zero in the entrained phase but greater than zero in the unentrained phase \cite{parker89}. For a given lattice, the onset of chaos occurs at the same $K$ as the appearance of single phase slips (Fig.~\ref{fig:qpsk}). The entrainment transition usually happens at the same $K$, although sometimes lower. 

\begin{figure}
\centering
\includegraphics[width=3.5 in]{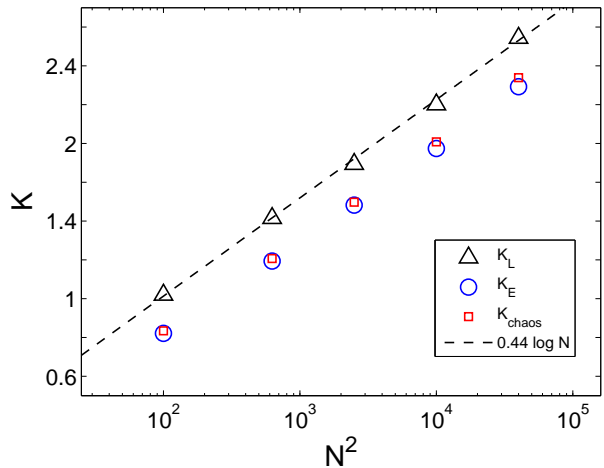}
\caption{\label{fig:kl} Coupling threshold for phase-locking $K_L$ (black triangles), entrainment $K_E$ (blue circles), and onset of chaos and single phase slips (red squares) for lattices with $N^2$ oscillators. Each data point is averaged over 50 disorder realizations. The standard deviation of the mean is about 0.02 for all points. Both $K_L$ and $K_E$ scale as $\log N$. The dashed line plots $0.44\log N$. Here, $K_E$ is defined as when a cluster encompasses half the lattice.}
\end{figure}

Averaging over the disorder, Fig.~\ref{fig:kl} shows that $K_E\sim \log N$, where $K_E$ is defined as when the largest cluster encompasses half the lattice. This scaling agrees with Ref. \cite{hong05}. The figure also shows that the onset of chaos and single phase slips coincides well with the entrainment transition.

\begin{figure}
\centering
\includegraphics[width=3.5 in]{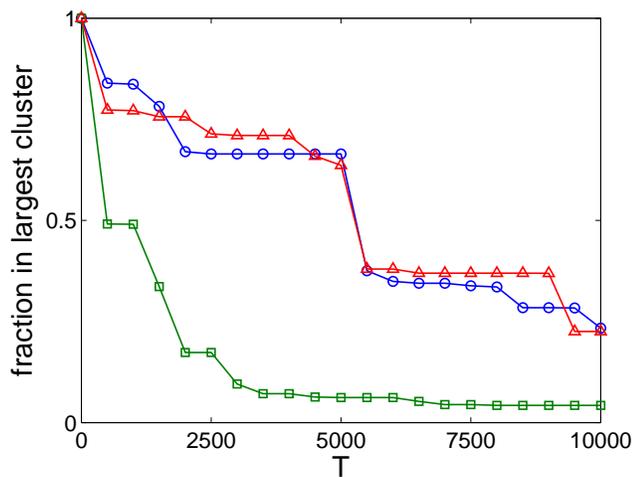}
\caption{\label{fig:qtavg} Relative size of the largest cluster vs.\ averaging time $T$ for three different $50\times 50$ disorder realizations. $K$ is below $K_E$, so the lattices are in the unentrained phase. The downward jumps are due to occasional wayward vortices that split up the cluster. This also shows that it is necessary for $T$ to be large in order to accurately distinguish the entrained and unentrained phases.}
\end{figure}

These observations indicate that the system-wide detrainment is caused by the random propagation of single phase slips. Indeed, Fig.~\ref{fig:qtavg} shows that when $K<K_E$, the largest cluster shrinks over time due to occasional phase slips within it, cutting it up until there is no longer a macroscopic cluster. 

\subsection{\label{sec:level3b}Vortices}

It is useful to speak of vortices to describe the propagation of phase slips. Looking at the numerical evolution of the solution, one sees that the phase is mostly smooth except for small regions where the phase gradient is large [Fig.~\ref{fig:lattice}(c)]. These are vortices, which are topological defects arising from the $2\pi$-periodicity of each $\theta$. The phase winds by $2\pi$ around a vortex. To identify vortices, one computes the lattice curl of the phase gradient around each unit cell of four oscillators \cite{kawamura97}. The curl is equal to the sum of the directed phase differences around a cell, where the phase differences have been shifted mod $2\pi$ into the range $(-\pi,\pi]$. The curl can be $+2\pi$, $-2\pi$, or 0, corresponding to a $+$ vortex, a $-$ vortex, or no vortex [Fig.~\ref{fig:lattice}(d)]. Note that a vortex exists on a unit cell of four oscillators, not just on one oscillator. 

From the above definition of a vortex on a discrete lattice, it follows that there must be an equal number of $+$ and $-$ vortices in the case of periodic boundary conditions. Also, the curl on a unit cell changes only when one of its $\Delta\theta$ crosses $\pi$ (mod $2\pi$). Thus, a vortex moves to a neighboring cell depending on which $\Delta\theta$ slips. In other words, vortex motion is equivalent to phase-slip propagation. Lastly, vortices are created when a pair of oscillators without neighboring vortices slips, resulting in two vortices of opposite polarity, one on either side. There are certain places in a lattice that tend to create vortices.

Since a vortex is topological, it exists until it meets and annihilates with a vortex of opposite polarity. Vortex paths delineate the boundaries of frequency clusters [Figs.~\ref{fig:lattice}(a) and \ref{fig:lattice}(b)]. When a vortex passes between a pair of oscillators, it causes a $2\pi$ phase slip. The accumulation of phase slips, due to repeated vortex crossings, leads to differences in average frequency.

Thus the entrained phase is characterized by vortices moving in fixed paths around the clusters, meaning that the vortices are locally confined and the cluster boundaries are time-independent. In the unentrained phase, vortices move inconsistently and sometimes wander off deeply into the formerly macroscopic cluster and chop it up (Fig.~\ref{fig:qtavg}). A signature of inconsistent motion is the presence of single phase slips during the averaging time, due to vortices that passed by only once, as opposed to regularly [Fig.~\ref{fig:dwavg}(d)]. This irregularity reflects the chaotic nature of the unentrained phase. 

In the rest of the paper, we call a cluster \emph{stable} when its vortices move consistently and \emph{unstable} when they move inconsistently. Instability connotes the presence of single phase slips and chaos. The fact that the entrainment transition occurs at a slightly lower $K$ than the onset of instability (Fig.~\ref{fig:kl}) indicates that the lattice becomes unentrained \emph{because} the clusters become unstable.

\subsection{\label{sec:level3c}Two types of entrainment transitions}

As $K$ decreases below $K_E$, the lattice transitions from the entrained to the unentrained phase. Simulations indicate that small lattices ($N\leq 50$) have a different pathway to the unentrained phase than large lattices ($N\geq 100$). The difference between the two is in the nature of the first microscopic cluster that appears at the locking threshold $K_L$. In small lattices, that cluster is usually made by a vortex pair, while in large lattices, it is usually made by a single vortex. Below, we describe the two types of clusters and their corresponding entrainment transitions.

\subsubsection{\label{sec:level3c1}Vortex-pair cluster}

The first type of cluster is made by a pair of oppositely-charged vortices that are created at a certain spot, travel along the cluster boundary, and then annihilate with each other. This cycle repeats periodically over time, so that the cluster has a different average frequency from its neighbor. We call this a \emph{vortex-pair cluster}, and an example is in Fig.~\ref{fig:lattice}.

As $K$ decreases, the shape of the cluster stays the same, because the vortices travel along the same paths although the vortices are produced more and more frequently. When $K<K_E$, the cluster is unstable and produces vortices which sometimes deviate from the original path. The liberated vortices sometimes cut across the macroscopic cluster so that the system becomes unentrained. Since a vortex exists until it annihilates with an oppositely charged one, a deviant vortex may travel a long distance before annihilating.

\begin{figure}
\centering
\includegraphics[width=3.5 in]{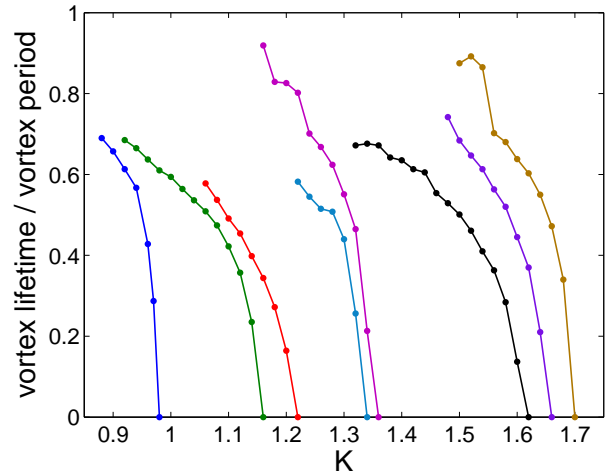}
\caption{\label{fig:tpstper} Ratio of vortex lifetime to the period of vortex production for the cluster formed at $K_L$. Each line corresponds to a disorder realization of size $N\times N$ and with a cluster containing $m$ oscillators. From left to right, $(N,m)$ is (10,3), (10,1), (10,1), (25,64), (25,60), (50,14), (25,6), and (50,260). The ratio is 0 when the cluster is formed at $K_L$ and increases as $K$ decreases until the ratio is on the order of 1. At that point, the cluster is unstable, and the vortices are sometimes liberated from their orbit around the cluster, leading to system-wide detrainment.}
\end{figure}

The stability of a vortex-pair cluster (and thus the existence of the entrained phase) is determined by the balance of two time scales: the lifetime of a vortex pair ($t_{life}$) and the period of pair production ($t_{per}$). When the cluster is formed at $K_L$, $t_{life}\ll t_{per}$. As $K$ decreases, $t_{life}$ increases while $t_{per}$ decreases (Fig.~\ref{fig:tpstper}). The cluster becomes unstable when $t_{life}\approx t_{per}$. In fact, a cluster with $t_{life}>t_{per}$ is never stable. This means that the entrainment transition happens when a new pair is produced immediately after the previous pair annihilates. So it seems that the transition is due to the balance of time scales, instead of the interaction between vortices of different clusters. We discuss this case further in Sec. \ref{sec:level4}.


\subsubsection{\label{sec:level3c2}Single-vortex cluster}

\begin{figure}
\centering
\includegraphics[width=3.5 in, viewport=40 80 320 360, clip]{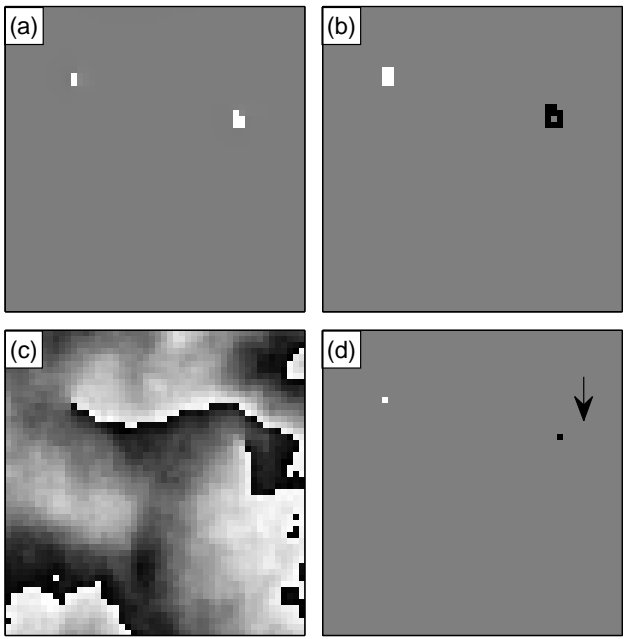}
\caption{\label{fig:phasevor_singlevor}Example of a $100\times 100$ lattice with two single-vortex clusters. Only a portion of the lattice is seen here. (a) Average frequency, showing the two single-vortex clusters and the macroscopic cluster. (b) Time-averaged vorticity, where black and white are opposite polarities. The vortices are each pinned to a cluster. (c) Phase (mod $2\pi$) at a point in time, where black is 0 and white is $2\pi$. (d) The equivalent lattice of vortices at that time, where black and white are opposite polarities. The arrow points to the source of the vortices, which does not produce any more vortices.}
\end{figure}

Another type of cluster is made by a single vortex that continually orbits it. We call this a \emph{single-vortex cluster}. It is usually composed of only one or a few oscillators. In the case of periodic boundary conditions, topological constraints require that there be an even number of vortices, so there are an even number of single-vortex clusters  (Fig.~\ref{fig:phasevor_singlevor}).

The stability of single-vortex clusters is an interesting phenomenon that invokes the topological nature of a vortex. Suppose $K>K_L$, so that all the oscillators are phase-locked. Due to the disorder of intrinsic frequencies, there is a time-independent phase gradient across the lattice. As $K$ decreases, the phase differences between neighbors increase. When $K=K_L$, there is a pair of oscillators on the verge of slipping relative to each other, and their phase difference is the largest in the lattice. When $K$ decreases below $K_L$, that pair does slip, producing two vortices that move. But instead of meeting and annihilating, the vortices are each pinned to a single-vortex cluster. A vortex has a phase field around it that accumulates by $2\pi$. The oscillator pair that produced them is actually prevented from slipping again by the phase field of the vortices: a vortex is positioned relative to the oscillator pair, so that its accumulating phase field balances the tendency of the pair to slip. Thus no more vortices are produced, and the cluster configuration is stable.

As $K$ decreases further with the vortices in this configuration, the phase difference of the oscillator pair increases again. When $K$ is low enough, it finally slips and produces more vortices. If the vortices do not find a new configuration to stop the oscillator pair, it will continue to produce vortices that detrain the lattice. Therefore, whether the lattice is entrained depends on whether the source of vortices is quenched. When it is quenched, the vortices are locally confined to single-vortex clusters, but when it is not quenched, the vortices wander throughout the system.

\section{\label{sec:level4}Stability of vortex-pair cluster}

Having described the features of the 2D model, we now do some analytical calculations. In particular, we are interested in how the entrainment threshold $K_E$ scales with the system size $N$. Since, a lattice becomes unentrained when the microscopic cluster formed at the locking threshold $K_L$ becomes unstable, we study the onset of instability. Here, we calculate $K_E(N)$ for small $N$, where the entrainment transition is determined by a vortex-pair cluster. We leave the stability of single-vortex clusters (and hence $K_E(N)$ for large $N$) for future work.

Our approach is based on the fact that the unentrained phase is chaotic while the entrained phase is not. Consider the interaction of the microscopic cluster with the macroscopic cluster that surrounds it. When the two clusters are stable, each may be considered as a single effective oscillator, since the constituent oscillators within each behave coherently. A system of two coupled oscillators is not chaotic, which is consistent with the assumption that the clusters are stable. However, when the clusters stop behaving as effective oscillators, the new degrees of freedom within each cause the system to become chaotic. In the presence of chaos, the vortices are no longer confined and proliferate to cause system-wide detrainment.

So the question is: when does a cluster stop behaving as an effective oscillator? Recall that a vortex-pair cluster is made by periodically produced vortex pairs. There are two time scales involved: $t_{life}$ is the lifetime of the vortex pair and $t_{per}$ is the period of pair production. In other words, $t_{life}$ is how long it takes for a pair to annihilate, and $t_{per}$ is the time between the creation of successive pairs. $t_{life}$ is also the duration of the cluster-wide phase slip. By definition, an effective oscillator can have only one phase slip during $t_{per}$. Thus, a cluster cannot behave as an effective oscillator when $t_{life}>t_{per}$. Thus the stability condition is
\begin{eqnarray} \label{eq:stable_cond}
t_{life}\ll t_{per}\;,
\end{eqnarray}
When this is satisfied, the clusters behave as effective oscillators \cite{note1}. We expect this condition to be satisfied for large but not small $K$. Note that this explanation is consistent with simulation results (Fig.~\ref{fig:tpstper}).


We proceed with a self-consistent argument: assuming that Eq.~\eqref{eq:stable_cond} is satisfied, we calculate $t_{life}$ and $t_{per}$ and then see when the condition is no longer satisfied. We consider the interaction between the first microscopic cluster and the macroscopic cluster that surrounds it. Simulations indicate that as $K$ decreases, the shape of the first cluster generally remains unchanged. Hence, we assume here that the first cluster keeps the same shape for $K>K_E$.

Let the first microscopic cluster and the macroscopic cluster be denoted by $A$ and $B$, respectively. We coarse-grain the $m_A$ oscillators of cluster $A$ into a single effective oscillator with phase $\theta^A=\frac{1}{m_A}\sum_{i\in A}\theta_i$ and intrinsic frequency $\omega^A=\frac{1}{m_A}\sum_{i\in A}\omega_i$. Let $\delta\theta^A$ describe an oscillator's deviation from the cluster phase: $\theta_i=\theta^A+\delta\theta^A_i$. We define similar quantities for cluster $B$. Since Eq.~\eqref{eq:stable_cond} is assumed to be satisfied, $\delta\theta^A_i$ and $\delta\theta^B_i$ can be taken to be time-independent. In other words, the phases are rigid within each cluster.

The phase difference $\phi=\theta^A-\theta^B$ satisfies
\begin{eqnarray}
\dot{\phi} & = & \Delta\omega - \frac{K}{\mu}\sum_{\langle ij\rangle}\sin(\phi+\delta\theta^A_i-\delta\theta^B_j)\;,
\end{eqnarray}
where $\Delta\omega = \omega_A - \omega_B$ and $\mu=\frac{m_Am_B}{m_A+m_B}$. The sum runs over the lattice edges that connect $A$ and $B$, since the coupling terms within each cluster cancel due to action-reaction symmetry. Let $q$ be the number of edges between $A$ and $B$. Then we write the sum of sines as a single sine:
\begin{eqnarray}
\dot{\phi} & = & \Delta\omega - \frac{\alpha K}{\mu}\sin(\phi+\beta)\;,
\end{eqnarray}
where
\begin{eqnarray}\label{eq:alpha}
\alpha(K)^2 & = &  q + \sum_{\langle ij\rangle\neq\langle kl\rangle}\cos((\delta\theta^A_i-\delta\theta^B_j)-(\delta\theta^A_k-\delta\theta^B_l)) \;,\nonumber\\ 
\end{eqnarray}
and $\beta(K)$ is a phase offset that does not matter in the following discussion.
The deviations $\delta\theta_i$ depend on $K$ and can be estimated using the linear solution (Sec. \ref{sec:level6b1}). In the limit of large $K$, all $\delta\theta_i\approx 0$: the coupling terms add coherently and $\alpha=q$. When $K$ is small, $\delta\theta_i$ is large: the sum in Eq.~\eqref{eq:alpha} is over $q(q-1)$ random numbers so $\alpha\sim\sqrt{q}$.

The period of vortex production $t_{per}$ is the period of $\phi$, since each vortex pair passing between $A$ and $B$ contributes $2\pi$ to $\phi$:
\begin{eqnarray}
t_{per} & = & \frac{1}{|\Delta\omega|} \int_0^{2\pi} \frac{d\phi}{1 - r\sin(\phi+\beta)}\;, \label{eq:t_per_int}\\
 & = & \frac{2\pi}{|\Delta\omega|\sqrt{1-r^2}} \;,
\end{eqnarray}
where $r=\frac{\alpha K}{\mu|\Delta\omega|}$. When $t_{per}$ diverges, cluster $A$ is synchronized with cluster $B$ and the lattice is phase-locked. Thus we identify $\alpha_L K_L=\mu\Delta\omega$. Assuming that $\alpha\approx\alpha_L$ in the range of $K$ that we are interested in,
\begin{eqnarray}\label{eq:tperscale}
t_{per} & = & \frac{2\pi \mu}{\alpha_L\sqrt{K_L^2-K^2}} \;.
\end{eqnarray}
Note that when $K\approx K_L$, $t_{per}\sim 1/\sqrt{K_L-K}$ as in a one-dimensional ring \cite{zheng98}.

We now estimate the lifetime of a vortex pair $t_{life}$. The presence of a vortex pair is a manifestation of the fact that the clusters are experiencing a phase slip relative to each other. Hence, we should calculate how long it takes the clusters to slip. First, we calculate the duration of a phase slip $t_{ps}$ in the effective-oscillator model. This is not the same as $t_{life}$, since vortices are not present in the effective-oscillator model. We will later add in the vortices by accounting for the fact that each cluster is spatially distributed. To estimate $t_{ps}$, we calculate the duration of time when $|\dot{\phi}|$ is large:
\begin{eqnarray}
t_{ps} & = & \frac{1}{|\Delta\omega|} \int_{\pi-\beta}^{2\pi-\beta} \frac{d\phi}{1 - r\sin(\phi+\beta)}\;, \\
 & = & \frac{\pi-2\tan^{-1} \frac{r}{\sqrt{1-r^2}}}{|\Delta\omega|\sqrt{1-r^2}}
\end{eqnarray}
For the sake of scaling, the exact limits on the integral do not matter. As $r$ decreases from 1 to 0, $t_{ps}$ increases slightly from $\frac{2}{|\Delta\omega|}$ to $\frac{\pi}{|\Delta\omega|}$. Thus,
\begin{eqnarray}\label{eq:tpsscale}
t_{ps}\sim\frac{1}{|\Delta\omega|}=\frac{\mu}{\alpha_L K_L}
\end{eqnarray}

To convert $t_{ps}$ to $t_{life}$, we use the fact that a vortex pair is created when the first edge phase difference between $A$ and $B$ crosses $\pi$ (mod $2\pi$) and annihilates when the last edge phase difference crosses $\pi$ (Sec. \ref{sec:level3b}).
Thus we consider
\begin{eqnarray}
\gamma(K) &=& \max_{\langle ij\rangle}(\delta\theta^A_i-\delta\theta^B_j) - \min_{\langle ij\rangle}(\delta\theta^A_i-\delta\theta^B_j),
\end{eqnarray}
where each term corresponds to one of the edges between $A$ and $B$. When a vortex pair is produced, $\gamma$ is how far $\phi$ needs to go before the pair annihilates. Due to the disorder of intrinsic frequencies, $\gamma$ increases as $K$ decreases. Since $\langle{\dot{\phi}}\rangle\sim \frac{1}{t_{ps}}$ during the lifetime of a vortex pair,
\begin{eqnarray}\label{eq:tlife}
t_{life} &\approx& \gamma(K)\,t_{ps}\sim  \frac{\gamma(K)\,\mu}{\alpha_L K_L}
\end{eqnarray}



Fortunately, it is not necessary to calculate $\gamma(K)$ explicitly. Since $K_E$ is defined as when $t_{life}\approx t_{per}$, we know that $\gamma(K_E) \approx 2\pi$, because then a vortex pair is created immediately after the previous one annihilates. We find $K_E$ by equating Eqs.~\eqref{eq:tperscale} and \eqref{eq:tlife} and arrive at the scaling
\begin{eqnarray}\label{eq:kescale}
K_E \sim K_L\;.
\end{eqnarray}
Thus, the entrainment transition is tied to the phase-locking transition. In Sec. \ref{sec:level5}, we show that $K_L\sim\log N$. This means that $K_E\sim\log N$, in good agreement with Fig.~\ref{fig:kl}.

Note that our argument only applies to lattices where the first microscopic cluster is a vortex-pair cluster ($N\leq 50$). We have not analytically studied the stability of single-vortex clusters, but Fig.~\ref{fig:kl} indicates that Eq.~\eqref{eq:kescale} would also apply to those cases.

\section{\label{sec:level5}Phase-locking}

For a given lattice of oscillators, when $K>K_L$, the system is phase-locked and all the oscillators have the same frequency. In this section, we calculate how $K_L$ depends on system size $N$. Since $K_L$ is different for different realizations of the intrinsic frequencies, we are actually interested in how the disorder-averaged $\langle K_L\rangle$ depends on $N$. We first consider 1D and then 2D. Although phase-locking in 1D has already been solved \cite{strogatz88a, strogatz88b}, we review it and then redo it using a linear approach in order to tackle 2D, where the usual approach does not work.

\subsection{1D}

\subsubsection{Usual approach}
Consider a one-dimensional chain of $N$ oscillators with open boundary conditions:
\begin{eqnarray}\label{eq:eom1d}
\dot{\theta}_1&=&\omega_1+ K\sin(\theta_{2}-\theta_1) \nonumber\\
\dot{\theta}_i&=&\omega_i+ K[\sin(\theta_{i-1}-\theta_i)+\sin(\theta_{i+1}-\theta_i)] \nonumber\\
&& \quad\quad\quad\quad\quad\quad i=2,\ldots,N-1 \\
\dot{\theta}_N&=&\omega_N+ K\sin(\theta_{N-1}-\theta_N) \nonumber
\end{eqnarray}
We assume, without loss of generality, that the average $\omega$ is 0. Then the phase-locked solution is given by ${\theta_i}$, such that all $\dot{\theta_i}=0$. Due to the open boundary conditions and action-reaction symmetry, one can solve for $\Delta\theta_i\equiv \theta_{i+1}-\theta_i$ by adding up the first $i$ equations:
\begin{eqnarray}
K\sin\Delta\theta_i&=&\sum_{j=1}^i \omega_j \quad\quad 1\leq i<N \;.
\label{eq:1d_soln}
\end{eqnarray}
A necessary and sufficient condition for the existence of a phase-locked solution is that:
\begin{eqnarray}
\label{eq:cond1d}
\max_i\,\{|\sum_{j=1}^i \omega_j|\}\leq K \;.
\end{eqnarray}
Thus, $K_L=\max\,\{|\sum_{j=1}^i \omega_j|\}$. There are $2^{N-1}$ solutions of Eq.~\eqref{eq:1d_soln}, since each $\Delta\theta_i$ can have two values. 

The unique stable solution is given by all $\cos\Delta\theta_i>0$. One can show this by considering the Jacobian $J_{ij}=\frac{d\dot{\theta}_i}{d\theta_j}$ at the phase-locked solution and requiring $\sum_{ij}x_i J_{ij}x_j<0$ for all perturbations $\{x_i\}$, such that $||\textbf{x}||>0$ and $\sum_{i=1}^n x_i=0$. The last condition is due to the fact that there is always a 0 eigenvalue of $J$ with eigenvector $(1,1,\cdots,1)^T$, corresponding to the uniform displacement of all phases. One finds:
\begin{eqnarray}
\sum_{ij}x_i J_{ij}x_j &=& -K\sum_{1\leq i<N}(x_i-x_{i+1})^2 \cos\Delta\theta_i\;.
\end{eqnarray}
Due to the arbitrariness of $\{x_i\}$, a necessary and sufficient condition for stability is that all $\cos\Delta\theta_i>0$. So only one of the $2^{N-1}$ phase-locked solutions is stable.

Using Eq.~\eqref{eq:cond1d}, it is possible to derive the scaling \mbox{$K_L\sim\sqrt{N}$} in 1D based on random-walk arguments \cite{strogatz88b}.

\subsubsection{Linear approach}
Consider the linearized version of Eqs.~\eqref{eq:eom1d}:
\begin{eqnarray}\label{eq:eom1dlin}
\dot{\theta}_1&=&\omega_1+ K(\theta_{2}-\theta_1) \nonumber\\
\dot{\theta}_i&=&\omega_i+ K[(\theta_{i-1}-\theta_i)+(\theta_{i+1}-\theta_i)] \nonumber\\
&& \quad\quad\quad\quad\quad\quad i=2,\ldots,N-1 \\
\dot{\theta}_N&=&\omega_N+ K(\theta_{N-1}-\theta_N) \nonumber
\end{eqnarray}
The solution to the linear model is straightfoward to find. Let the steady state of the linear model be $\{\theta_i^{lin}\}$, which always exists. One may get the phase-locked solution of the corresponding nonlinear model by making the ansatz:
\begin{eqnarray}
\label{eq:lintononlin}
\Delta\theta_i &=& \sin^{-1}\Delta\theta_i^{lin} \;.
\end{eqnarray}
Plugging this into Eqs.~\eqref{eq:eom1d}, one immediately returns to Eqs.~\eqref{eq:eom1dlin}, which are all zero in the steady state. Thus, it is easy to go from the linear solution to the nonlinear phase-locked state. In light of this, a necessary and sufficient condition for phase-locking in the nonlinear model is
\begin{eqnarray}
\max \{|\Delta\theta_i^{lin}|\}\leq 1 \;,
\end{eqnarray}
which is equivalent to Eq.~\eqref{eq:cond1d}.

This approach has a nice intuitive interpretation. The coupling force is stronger in the linear model, since $\Delta\theta$ is steeper than $\sin\Delta\theta$. Imagine starting from the linear steady state and replacing a single linear term in Eqs.~\eqref{eq:eom1dlin} with the original nonlinear term. In order to get the same amount of coupling force to maintain $\dot{\theta}=0$, the corresponding $\Delta\theta$ must be increased in magnitude. This is seen in Eq.~\eqref{eq:lintononlin}. However, if $|\Delta\theta^{lin}|>1$, it is impossible to get the same amount of coupling force, so there is no equivalent phase-locked solution.

Consider the dynamics. Suppose the nonlinear model starts with all $\theta_i=0$. Since the phase differences are small, the dynamics are initially linear, and the phases approach the linear steady state. If $|\Delta\theta_i^{lin}|\leq 1$, $\Delta\theta_i$ converges to $\sin^{-1}\Delta\theta_i^{lin}$, but if $|\Delta\theta_i^{lin}|>1$, $\Delta\theta_i$ cannot converge so it continues to increase and phase slips. After the phase slip, all phase differences are small again ($\mbox{mod }2\pi$), so the dynamics are linear for sometime until $\Delta\theta_i$ slips again. Thus, if $K$ is large so that all $|\Delta\theta^{lin}|\leq 1$, the system is phase-locked. If one $|\Delta\theta^{lin}|> 1$, there will be a break in the chain there. If more than one $|\Delta\theta^{lin}|> 1$ and their density is small, each one corresponds to a break.

Thus, we are doing a local self-consistency check on the linear solution. If the linear solution has a big phase gradient somewhere, it is locally inconsistent there and will phase slip. We also note that this is an easy way to find the phase-locked solution in 1D with periodic boundary conditions.

\subsection{2D}

\subsubsection{\label{sec:level6b1}Phase-locked solution}
In 2D, the usual approach does not work for finding the phase-locked solution, either with open or periodic boundary conditions. This is because the connectivity prevents one from isolating a single sine term like in Eq.~\eqref{eq:1d_soln}. However, the linear approach is still applicable in 2D, since the linear solution, Eq.~\eqref{eq:2dss}, always exists. In the next section, we use this approach to show that $K_L\sim\log N$.


\begin{figure*}
\centering
\includegraphics[width=7 in]{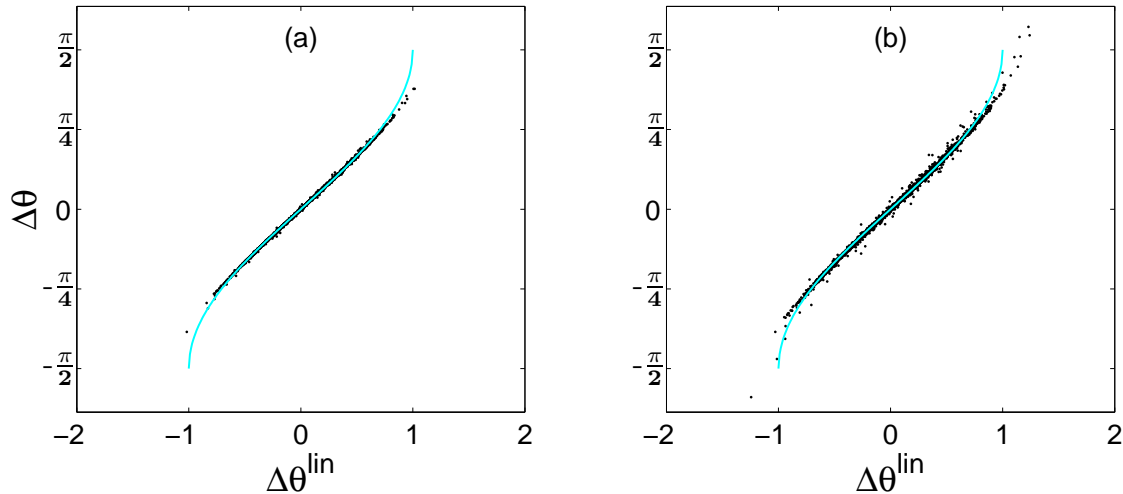}
\caption{\label{fig:dthlin} The actual phase-locked solution $\Delta\theta$ vs.\ the linear steady-state solution $\Delta\theta^{lin}$ for (a) $K=1.2K_L$ and (b) $K=K_L$. There is an approximate inverse-sine relationship (cyan line). In (b), the phase-locked solution sometimes has $|\Delta\theta|>\frac{\pi}{2}$ due to the presence of loops in 2D. Both plots are based on the same $50\times 50$ lattice.}
\end{figure*}

There is a catch though in 2D due to the presence of loops. Suppose one has the linear steady-state solution $\{\theta_{ij}^{lin}\}$. The linear approach says that $\Delta\theta=\sin^{-1}\Delta\theta^{lin}$ for each horizontal and vertical pair of oscillators. This is actually not a physically allowed solution since the sum of $\Delta\theta$ around a unit cell would generally not be a multiple of $2\pi$. This was not a problem in 1D because there were no loops. Thus in 2D, the linear approach is no longer exact. However, Fig.~\ref{fig:dthlin} shows that the relationship between the actual $\Delta\theta$ and $\Delta\theta^{lin}$ is still approximately an inverse sine.

There is another difference in 2D due to loops: it is possible to have a stable phase-locked solution with $\cos\Delta\theta<0$. Consider the Jacobian of Eqs.~\eqref{eq:eom2d}: $J_{ijkl}=\frac{d\dot{\theta}_{ij}}{d\theta_{kl}}$. The stability condition is that $\sum_{ijkl}x_{ij}J_{ijkl}x_{kl}<0$ for all perturbations $\{x_{ij}\}$, such that $||\textbf{x}||>0$ and $\sum_{ij}x_{ij}=0$. One finds
\begin{eqnarray}
\textbf{xJx}&=&-K\sum_{ij}[(x_{ij}-x_{i-1j})^2\cos(\theta_{i-1j}-\theta_{ij}) \nonumber\\
&&\quad\quad+(x_{ij}-x_{ij-1})^2\cos(\theta_{ij-1}-\theta_{ij})] \;.
\end{eqnarray}
One might think that a necessary condition for stability is that all $\cos\Delta\theta>0$, since if a $\cos(\theta_{i-1j}-\theta_{ij})<0$ and all $x$-differences but $x_{ij}-x_{i-1j}$ were 0, then $\textbf{xJx}>0$. The catch is that one cannot set all but one $x$-difference to 0; due to the loop nature there must be at least one other nonzero $x$-difference to compensate. Thus the requirement that all $\cos\Delta\theta>0$ is only a sufficient condition for stability. Indeed, for $K\approx K_L$, the stable phase-locked solution has some $\cos\Delta\theta<0$ [Fig.~\ref{fig:dthlin}(b)]. 

Note that these are features not just of the 2D lattice, but of any oscillator network with loops. In 2D, the core issue is that there are $N^2$ phases but $2N^2$ phase differences, so specifying the differences overspecifies the system.

\subsubsection{Phase-locking criterion} \label{sec:plcriterion}

In 1D, there was phase locking if and only if $\max\,|\Delta\theta^{lin}|\leq 1$. This raises the question whether there is also a critical value $\Delta\theta^{lin}_c$ for $\max\,|\Delta\theta^{lin}|$ in 2D. The critical value is probably not an absolute constant since the linear approach is an approximation in 2D. According to simulations, the critical value is actually narrowly distributed around 1.27 and seems to be independent of system size (Table \ref{tab:dthlinc}). The fact that the value is fairly consistent means that the intuition behind the linear approach is still valid: if the linear solution has small phase gradients everywhere, then there is phase-locking, but if it has a big gradient somewhere, that place will be a source of vortices. Thus, the linear solution predicts where the first vortices are formed.

\begin{table}
\caption{\label{tab:dthlinc}Critical values of phase differences of the linear steady-state solution for $N\times N$ lattices. If $\max\,|\Delta\theta^{lin}|<\Delta\theta^{lin}_c$, the system is phase-locked; otherwise, there are vortex sources. These values were found numerically for 50 realizations for each size. $\Delta\theta^{lin}_c$ is randomly distributed with the sample standard deviation given by the digit in parantheses.}
\begin{ruledtabular}
\begin{tabular}{ccc|cc}
&$N$ && $\Delta\theta^{lin}_c$ &\\
\hline
&10 && 1.28(7)&\\
&25 && 1.28(6)&\\
&50 && 1.26(5)&\\
&100 && 1.27(5)&\\
&200 && 1.25(5)&\\
\end{tabular}
\end{ruledtabular}
\end{table}

This provides a way to calculate the critical coupling for phase-locking $K_L$, since $\Delta\theta^{lin}$ is inversely related to $K$ in Eq.~\eqref{eq:2dss}. This reasoning may seem circular, since we first found $\Delta\theta^{lin}_c$ by numerically checking for phase-locking and are now using it to determine the phase-locking criterion. Actually, one can derive the scaling $K_L(N)$ just by positing the existence of some $\Delta\theta^{lin}_c$. The actual value for $\Delta\theta^{lin}_c$ is only in the prefactor, as seen in Eq.~\eqref{eq:kllogn}.

We first calculate the probability of phase-locking $P(K,N)$ for a given coupling $K$ and system size $N$. The condition for phase-locking is that all $|\Delta\theta^{lin}|<\Delta\theta^{lin}_c$. Since $\Delta\theta^{lin}$ is a random variable with standard deviation $\sigma_{\Delta\theta^{lin}}$ given in Eq.~\eqref{eq:sddthlin}, $P(K,N)$ is the probability that all $2N^2$ phase differences are less than the threshold. Assuming that the $\Delta\theta^{lin}$ are independent,
\begin{eqnarray}
P(K,N)&=& \left[\mbox{erf}\left(\Delta\theta^{lin}_c K\sqrt{\frac{2\pi}{\log N}}\right)\right]^{2N^2}
\end{eqnarray}
The distribution of $K_L$ is $\rho_{K_L}(K)=\frac{dP(K,N)}{dK}$. The disorder-averaged $K_L$ is estimated by the $K$ that maximizes $\rho_{K_L}$. One gets the implicit equation
\begin{eqnarray}
\left[\mbox{erf}\left(\Delta\theta^{lin}_c \langle K_L\rangle\sqrt{\frac{2\pi}{\log N}}\right)\right]\langle K_L\rangle= \qquad\qquad\qquad\qquad\nonumber \\
\frac{2N^2-1}{\Delta\theta^{lin}_c }\sqrt{\frac{\log N}{8\pi}}\exp\left(-\frac{2\pi (\Delta\theta^{lin}_c)^2 \langle K_L\rangle^2}{\log N}\right)\;.\quad
\end{eqnarray}
Taking the logarithm of both sides and considering the largest terms in the limit of large $N$,
\begin{eqnarray}
\label{eq:kllogn}
\langle K_L\rangle \approx\frac{1}{\Delta\theta^{lin}_c\sqrt{\pi}}\log N \approx 0.44 \log N \;,
\end{eqnarray}
which agrees well with numerics (Fig.~\ref{fig:kl}). Thus, $K_L$ scales as $\log N$.

The scaling $K_L\sim\log N$ is consistent with bounds in a previous work \cite{strogatz88b}. A necessary condition for phase-locking is that all $|\omega_i|\leq 4K$. The probability of phase-locking is bigger than the probability that all $N^2$ $\omega$'s satisfy this, leading to a lower bound on $K_L$ that scales as $\sqrt{\log N}$. 


We also note that the above approach to calculate $K_L$ is appropriate when the initial phases are zero, since the dynamics are linear. When the initial phases are randomized, the system might not lock when $K>K_L$. The clusters in this case are due to vortices created by the initial conditions instead of spontaneously.

\section{\label{sec:level6}Conclusion}

In conclusion, we have studied the entrainment and phase-locking transitions in the 2D Kuramoto model. We derived the scaling $K_E\sim\log N$ for small lattices, in agreement with simulations. We have relied on the insight that in 2D, a system is unentrained when a cluster stops behaving as an effective oscillator, allowing vortices to be liberated.

The next step would be to analytically derive $K_E(N)$ for large lattices, since one would like to know whether it is finite in the limit $N\rightarrow\infty$. This would require explaining the stability of single-vortex clusters, which seems like a challenging theoretical problem. However, doing so would answer an open question regarding the Kuramoto model. It would also be an interesting problem to predict cluster structure using a real-space renormalization group approach, as was done in 1D \cite{kogan09,lee09}.

Finally, one should study the role of topological defects in other synchronization models, such as higher dimensions, complex networks, long-range interactions, or other oscillator types. For example, one can create a small-world network by randomly rewiring some links in a 2D lattice, so that the average distance between oscillators drops dramatically \cite{watts98}. One should see whether the unentrained phase is still characterized by inconsistent vortex motion, chaos, and time-dependent cluster structure. Thus, the results in this work may prove useful in building a solid mathematical understanding of synchronization in a variety of situations.

This work was supported by Boeing. GR thanks the Research Corporation and the Packard Foundation for their generous support.

\bibliography{draft}

\begin{thebibliography}{28}
\bibitem[{not()}]{note1}
\bibinfo{note}{To clarify, what matters is whether the microscopic cluster acts
  as an effective oscillator. Usually if a macroscopic cluster stops behaving
  as an effective oscillator, an additional microscopic cluster appears within
  it. This does not affect the stability of the first microscopic cluster if
  the new cluster is far away, so the modified macroscopic cluster returns to
  behaving as an effective oscillator.}
\end{thebibliography}

\end{document}